\documentclass[aps,prd,superscriptaddress,nofootinbib]{revtex4-1}%twocolumn
\usepackage{graphicx}
\usepackage{epsfig}
\usepackage{amsmath}
\usepackage{subfigure}
\usepackage{ulem}
\usepackage{color}
\usepackage{latexsym} % Gets \Box etc
\usepackage{amssymb}  % \gtrsim, \geqslant, etc etc: see amsguide.ps
\usepackage{amsfonts} % \mathfrak and \mathbb{x} (Blackboard bold)
\usepackage{amsbsy}   % \pmb and \boldsymbol
\usepackage{multirow} % for multirows in tables
\usepackage{pstricks}
\usepackage{slashed}
\usepackage{dcolumn}
\usepackage{xparse}
\usepackage{bm}
\usepackage{hyperref}
\newcommand{\al}{\alpha}
\newcommand{\be}{\beta}
\newcommand{\ga}{\gamma}

\newcommand{\De}{\Delta}
\newcommand{\ep}{\varepsilon}

\newcommand{\la}{\lambda}
\newcommand{\La}{\Lambda}

\newcommand{\si}{\sigma}
\newcommand{\Si}{\Sigma}
\renewcommand{\th}{\theta}   % LaTeX: \th already defined

\newcommand{\om}{\omega}
\newcommand{\Om}{\Omega}

\newcommand{\beq}{\begin{equation}}
\newcommand{\eeq}{\end{equation}}
\newcommand{\ba}{\begin{array}}
\newcommand{\ea}{\end{array}}
\newcommand{\bea}{\begin{eqnarray}}
\newcommand{\eea}{\end{eqnarray}}
\newcommand{\bi}{\begin{itemize}}    
\newcommand{\ei}{\end{itemize}}
\newcommand{\ben}{\begin{enumerate}} 
\newcommand{\een}{\end{enumerate}}
\newcommand{\bc}{\begin{center}}
\newcommand{\ec}{\end{center}}
\newcommand{\bl}{\begin{flushleft}}
\newcommand{\el}{\end{flushleft}}
\newcommand{\br}{\begin{flushright}}
\newcommand{\er}{\end{flushright}}

\newcommand{\nn}{\nonumber \\}
\newcommand\Eqn[1]{Eq.~(\ref{#1})}  % includes ``Eq.'' in front
\newcommand\Fig[1]{Fig.~(\ref{#1})} % includes ``Fig.'' in front
\newcommand{\mr}{\mathrm}
\newcommand{\mb}{\mathbf}

\newcommand{\mc}{\mathcal}
\newcommand{\mi}{\mathrm{i}}
\newcommand{\me}{\mathrm{e}}
\newcommand{\dif}{\mathrm{d}}
\newcommand{\p}{\partial}

\newcommand{\tr}{\hbox{tr}}
\newcommand{\Tr}{\hbox{Tr}}

\newcommand{\MeV}{{\rm MeV}}
\newcommand{\GeV}{{\rm GeV}}

\newcommand{\<}{\langle}
\renewcommand{\>}{\rangle}   % LaTeX: \> already defined
\renewcommand{\l}{\left}
\renewcommand{\r}{\right}

\newcommand\comment[1]{ \hbox{[{\it Comment suppressed here.}\/]} }
\newcommand\hide[1]{}
\newcommand{\skipover}[1]{}

\begin{document}


\title{Phase diagram of two-color QCD matter at finite baryon and axial isospin densities}

\author{Jingyi Chao}%
\email{jychao@impcas.ac.cn}
\affiliation{Institute of Modern Physics, Chinese Academy of Sciences, Lanzhou, 730000, China}
\date{\today}

%\address[a1]{Institute of Modern Physics, Chinese Academy of Sciences, Lanzhou, 730000, China}

\begin{abstract}
We study the two-color QCD matter within two fundamental quark flavors via both chiral perturbation theory and Nambu--Jona-Lasinio model methods. The effective Lagrangian described by low lying meson and baryon, i.e., diquark, is derived, where the excitations locate in the extended $\mr{SU}(4)$ flavor symmetry space. We determine the leading order terms on the dependence of the baryon and axial isospin densities. Then, the two-color NJL model is employed to run the numerical simulation and the phase diagram in the plane of $\mu-\nu_{5}$ is plotted.

\end{abstract}
% ======================================================================
%\pacs{ {13.40.Gp}, {12.38.-t}, {12.38.Lg}}% PACS, the Physics and Astronomy
%\keywords{charged pion condensate, electromagnetic fields, isospin chemical potential, quantum chromodynamics}axial isospin chemical potential \sep diquark condensate \sep spin-one paring \sep two-color QCD

\maketitle

%\bigskip
%\maketitle
% ======================================================================
\section{Introduction}
\label{int}
Color superconductivity of Quantum Chromodynamics (QCD) matters at high baryon density, such as in the core of the neutron star, has been extensively studied, seen the review in~\cite{RevModPhys.80.1455}. Lots of works have shown that the superconducting gaps is large enough to be observed in astrophysics and/or the laboratory of heavy-ion collisions. However, the first principle computations hasn't well approached due to the sign problem of the lattice Monte Carlo simulation for $N_{c}=3$~\cite{PhysRevLett.91.222001,PhysRevD.75.116003}. As a remaining theoretical challenge, it was proposed to understand the finite baryon density quark matter via other simpler QCD-like theories. For example, at $N_{c}=2$, the fermion determinant is guaranteed to be real at nonzero chemical potential due to the additional anti-unitary symmetries, allowing a lattice evaluation for both fundamental and adjoint quarks~\cite{PESKIN1980197,PhysRevLett.72.2531,HANDS1999327,KOGUT2000477,PhysRevD.64.016003,DUNNE2003307,MUROYA2003305,PhysRevD.70.054013,ALLES2006124,PhysRevD.74.014506,PhysRevD.85.074007}. Plus, the flavor space is extended from the usual $\mr{SU}(N_{f})$ to a larger $\mr{SU}(2N_{f})$ one~\cite{PESKIN1980197,KOGUT2000477,PhysRevD.64.016003}. $N_{c}=2$ theory is not exactly same as the full three-color QCD since its statistical freedom of baryon is Bose but not Fermi distribution. Nevertheless, it sheds a light in searching the qualitative mechanism of the phase transitions, the forming of the diquark condensation, critical phenomena and so on for reality QCD matter. 

Many interests have been paid to the investigation of the QCD phase diagram at not only finite baryon but also nonzero isospin densities~\cite{PhysRevD.64.016003,PhysRevLett.86.592,PhysRevD.66.014508,PhysRevD.68.014009}. The main reason is that the dense hadronic matter is isotropically asymmetric in heavy-ion collision experiments. The inside compact stars is expected to be isospin asymmetries, as well. A large class of effective theories, such as low energy effective theory~\cite{PhysRevLett.86.592,PhysRevD.64.016003}, Nambu--Jona-Lasinio (NJL) type models~\cite{TOUBLAN2003212,PhysRevD.69.096004,PhysRevD.71.116001,PhysRevD.79.034032}, linear sigma model~\cite{PhysRevD.78.014030}, Polyakov-Quark-Meson model~\cite{PhysRevD.88.074006,STIELE201472}, Dyson-Schwinger equation~\cite{PhysRevC.75.035201} and functional renormalization group methods~\cite{SVANES201116,KAMIKADO20131044} are adopted to explore the orientation of the vacuum expectation value (VEV) and the matter phase structure in this additional aspect.

The exploration along the way in an external electromagnetic (EM) environment~\cite{PhysRevD.39.3478,BABANSKY1998272,PhysRevLett.95.152002,PhysRevD.76.105030,PhysRevD.77.014021,PhysRevLett.105.042001} becomes gloomy since the charge separation along the magnetic field was observed in the non-central heavy ion collisions at the Relativistic Heavy Ion Collider (RHIC) ~\cite{PhysRevLett.103.251601}. In nature, the background EM field exists in various physical systems, such as the compact star
~\cite{Duncan1992}, the early universe~\cite{GRASSO2001163} and the laboratories of RHIC or Large Hadron Collider (LHC)~\cite{PhysRevLett.103.251601,PhysRevLett.110.012301,PhysRevC.85.044907}. EM field excites the charged particles both in hadronic and quark-gluon plasma (QGP) phases, which offers a unique paradise to study the features of QCD vacuum structure and the thermodynamics of strong interaction matter. Several approaches have been established to understand the appearing of magnetic catalysis at low temperature but inverse magnetic catalysis near $T_{c}$ for magnetized quark matter~\cite{MIRANSKY20151,PhysRevLett.73.3499,PhysRevD.86.071502,PhysRevD.88.054009,Bruckmann2013}. It was suggested that QCD vacuum will become a superconductor due to charged $\rho$ condensation~\cite{PhysRevD.82.085011}. Recently, the EM chiral anomaly has trigged a new interest to investigate the QCD phase diagram under the (anti) parallel EM environment, i.e., electric and magnetic fields are (anti) parallelized~\cite{CAO20161,PhysRevD.92.105030,PhysRevD.93.094021}. In the work of~\cite{CAO20161}, the authors found that the pion superfluid is energy favored with the increasing of the strength of the EM fields. Indeed, chiral anomaly is closely linked to the explanation of various quantum phenomena taking place in the EM fields, such as the chiral magnetic effect~\cite{PhysRevD.78.074033}, the chiral magnetic wave~\cite{PhysRevD.83.085007}, the chiral electrodynamics~\cite{PhysRevLett.110.232302}, etc.

Consequently, one has to pay attention to the behaviors of diquark in dense baryon matter since the
EM field is acting on all the charged quasi-particles and the diquark is becoming the degree of freedom in low energy effective theory of QCD matter at finite chemical potential, $\mu$~\cite{KOGUT2000477,DUNNE2003307}. If only QCD interaction is under considering, the axial isospin currents is anomaly free. Instead, the anomaly is arising while the quarks coupling to electromagnetism. The corresponding current is given by
\begin{equation}
	\p_{\mu}j_{5}^{\,\mu 3}=-\frac{e^{2}}{16\pi^{2}}\ep^{\al\be\mu\nu}F_{\al\be}F_{\mu\nu}\cdot\tr\l[\tau_{3}Q^{2}\r],
\end{equation}
where $Q$ is the matrix of quark electric charges, $\tau_{3}$ is the Pauli matrix in flavor space and $F$ is the field strength. The related process is the decay of a neutral pion into two photons~\cite{PhysRev.177.2426,Bell1969,PhysRevD.63.076010}. In order to qualitatively study this charge asymmetry in flavor space under the EM fields, we introduce an axial isospin chemical potential $\nu_{5}$ and then $\nu_{5}\bar{\psi}\ga_{0}\ga_{5}\tau_{3}\psi$ is built into the Lagrangian in quark sector~\cite{PhysRevD.94.116016}. 

To research the properties of diquark in the EM systems, in this work, we learn the QCD phase diagram in the plane of baryon, $\mu$, and axial isospin, $\nu_{5}$, chemical potentials in the framework of two-color QCD theory ($\mr{QC_{2}D}$), which constituted by an two-color gauge group with two Dirac flavors in the fundamental representation. It dynamically generates quasi-particles in terms of sigma, pions and baryons (diquarks). We write down the specific model in the section~\ref{qc2d}. In Sec.~\ref{su4}, we achieve the underlying effective Lagrangian terms. By the realized static low-energy effective $\mc{L}$, it allows us to identify the breaking pattern of global symmetries. The numerical results presenting by NJL model are shown in Sec.~\ref{omega}. We enclose the conclusions in the final section.

% ======================================================================
\section{Lagrangian in two-color QCD}
\label{qc2d}

We start with the following NJL-type Lagrangian formulated in $\mr{QC_{2}D}$ model, whose effective four-fermion interactions are acquired in a gluon inspired manner. The detail form is referred to the paper~\cite{PhysRevD.70.054013}, shown as
\begin{equation}
	\mc{L}_{\mr{NJL}}=\bar{\psi}\l(\mi\ga^{\mu}\p_{\mu}+\mu\ga_{0}+\nu_{5}\ga_{0}\ga_{5}\tau_{3}-m_{0}\r)\psi+\mc{L}_{\bar{q}q}+\mc{L}_{qq}+(\text{color triplet terms})
\end{equation}
\begin{equation}\label{eqn_L_qqbar}
	\mc{L}_{\bar{q}q}=\frac{G}{2}\l[(\bar{\psi}\psi)^2+(\bar{\psi}\mi\ga_{5}\vec{\tau}\psi)^2\r]
\end{equation}
\begin{equation}
	\mc{L}_{qq}=\frac{H}{2}(\bar{\psi}\mi\ga_{5}\tau_{2}t_{2}C\bar{\psi}^{T})({\psi}^{T}C\mi\ga_{5}\tau_{2}t_{2}{\psi})-\frac{H}{4}(\bar{\psi}\ga_{3}\tau_{1}t_{2}C\bar{\psi}^{T})({\psi}^{T}C\ga_{3}\tau_{1}t_{2}{\psi})
\end{equation}
where $C=\mi\ga_{0}\ga_{2}$ is charged conjugate operator, $m_{0}$ is the current quark mass, $\tau_{i}$ and $t_{i}$ are the Pauli matrices in flavor and color spaces, respectively. The two coupling constants $G$ and $H$ are connected by a Fierz transformation in color space. Turning out, $G=H$ at $N_{c}=2$. The operators $\tau_{2}, t_{2}$ are antisymmetric matrix acting in flavor and color space. 

Since $\nu_{5}=\mu_{L}^{u}-\mu_{R}^{u}=\mu_{R}^{d}-\mu_{L}^{d}$, the left handed $u$ quark is at the same Fermi surface with the right handed $d$ quark, while as the baryon density of right handed $u$ quark is equal to left handed $d$ quark, the corresponding LR bound states composed by the quarks which standing on the equal Fermi surfaces are $\psi_{L}^{u}\psi_{R}^{d}\pm\psi_{R}^{u}\psi_{L}^{d}$. Confining in the most attractive anti-symmetric color sector, the Dirac structure of total antisymmetric LR diquark $\psi_{L}^{u}\psi_{R}^{d}-\psi_{R}^{u}\psi_{L}^{d}$ is in the repulsive $(C\ga_{0}\ga_{5})$ channel with zero spin and isospin, which cannot form a bound particle. On the other hands, the flavor symmetric state, $\psi_{L}^{u}\psi_{R}^{d}+\psi_{R}^{u}\psi_{L}^{d}$ is attractive with Dirac operator $(C\ga_{3})$. Symmetric flavor diquark pairing patterns have been explored in the same or one flavor color superconductor~\cite{PhysRevD.62.094007,PhysRevD.67.054018,PhysRevLett.90.182002,PhysRevLett.91.242301,PhysRevD.72.034008,PhysRevD.75.054022,PhysRevD.82.085007}. It was suggested, excepted of the most attractive scalar diquark ($C\ga_{5}$), the axial spin one diquark pairs are formed. Obviously, at finite $\nu_{5}$, our system prefers to the polar phase ($s=1$, $s_{z}=0$) of symmetric spin and symmetric flavor $\tau_{1}$ direction due to the Pauli exclusion principle. 

In the works of\cite{CAO20161,Chao:2018ejd}, the authors figured out that the QCD vacuum takes a chiral rotation from scalar $\si$ channel to pseudo scalar $\pi_{i}$ channels once the EM fields turning on due to the chiral anomaly. Therefore, applying the Hubbard-Stratonovich transformation, we introduce three kinds of auxiliary meson and diquark fields in this work, where $\pi_{3}=-\frac{G}{2}\<\bar{\psi}\mi\ga_{5}\tau_{3}\psi\>$ representing for pseudo scalars, $\De=-\frac{H}{2}\<\psi^{T}\mi\ga_{5}\tau_{2}t_{2}C\psi\>$ and $\De^{*}=-\frac{H}{2}\<\bar{\psi}\mi\ga_{5}\tau_{2}t_{2}C\bar{\psi}^{T}\>$ denoting for complex scalar diqurks, $d=-\frac{H}{4}\<\psi^{T}\ga_{3}\tau_{1}t_{2}C\psi\>$ and  $d^{*}=-\frac{H}{4}\<\bar{\psi}\ga_{3}\tau_{1}t_{2}C\bar{\psi}^{T}\>$ regarding for complex axial vector diquarks. In terms of Nambu-Gorkov bispinors
\begin{equation}
    \psi=\frac{1}{\sqrt{2}}\l(\begin{matrix}
	q \\ C\bar{q}^{T}
	\end{matrix}\r),\quad \bar{\psi}=\frac{1}{\sqrt{2}}\l(\bar{q}, q^{T}C\r),
\end{equation}
the effective thermodynamical potential is formed as
\begin{equation}
	\Om_{\text{eff}}=\frac{1}{2}\bar{\psi}\mc{S}^{-1}\l(p;\pi_{3},\De,d\r)\psi-\frac{\pi_{3}^2+|\De|^2+|d|^2}{4G}.
\end{equation}
Here, the inverse propagator of fermion is
\begin{equation}
	\mc{S}^{-1}(p)=\l(\begin{matrix}
    \slashed{p}-M+\ga_{0}\mu+\ga_{0}\ga_{5}\tau_{3}\nu_{5} &\tilde{\De}\\
    -\tilde{\De}^{\dagger}&\slashed{p}-M^{T}-\ga_{0}\mu+\ga_{0}\ga_{5}\tau_{3}\nu_{5}
\end{matrix}\r),
\end{equation}
where $M=\l(m_{0}-\si\r)\tau_{0}-\mi\ga_{5}\tau_{3}\pi_{3}$ and $\tilde{\De}=\ga_{5}\tau_{2}\De+\ga_{3}\tau_{1}d$. The color index has been abbreviated since it becomes trivial in the further calculation. 
% ======================================================================
\section{Mesons and Diquarks in chiral perturbation theory} 
\label{su4}
In this section, we briefly explain the reasons why quasiparticles of $\pi_{3}$, $\De$ and $d$ are chosen from the view of chiral perturbation theory. 

The fundamental representation of group $\mr{SU}(2)$ is pseudo-real and isomorphic to its complex conjugate representation with the isometry given by $S=\mi\si_{2}$~\cite{PESKIN1980197,KOGUT2000477,PhysRevD.64.016003}.
This theory has an enlarge flavor symmetry $\mr{SU}(2N_{f})$ in the name of spinors
\begin{equation}
	\Psi=\l(\begin{matrix}
	\psi_{L} \\ \tilde{\psi}_{R}
	\end{matrix}\r),\quad \Psi^{\dagger}=\l(\psi_{L}^{\dagger},\tilde{\psi}_{R}^{\dagger}\r),
\end{equation}
 where $\tilde{\psi}_{R}=-\mi\si_{2}S\psi_{R}^{*}$ and $-\mi\si_{2}$ is the charge conjugation matrix $C$ for the $R$-chiral component. In $\mr{SU}(2N_{f})$ flavor space, it manifest its symmetry and chiral component at the same time, establishing a connection between quarks and antiquarks~\cite{PESKIN1980197,KOGUT2000477,PhysRevD.64.016003}. As a consequence, color singlet baryons are composed by two quarks and the scalar diquarks become degenerate with pseudo meson, pions.
 %For $N_{f}=2$, the enlarged flavor group, $\mr{SU}(4)$, is called Pauli-Guersey symmetry

Following above expression, the standard kinetic part of the Euclidean $\mr{QC_{2 }D}$ Lagrangian can be written as~\cite{PhysRevD.85.074007}
\begin{equation}
	\mc{L}_{\mr{kin}}=\Psi^{\dagger}\mi\si^{\mu}D_{\mu}\Psi,
\end{equation}
where Hermitian gamma matrix $\si^{\mu}=(-\mi,\vec{\si})$ and the covariant derivative is $D_{\mu}=\p_{\mu}+\mi A_{\mu}^{a}T_{a}$. The coupling is absorbed in the gauge fields. In terms of $\Psi$, quark mass term becomes~\cite{PhysRevD.85.074007}
\begin{equation}
	\mc{L}_{\mr{mass}}=\frac{m_{0}}{2}\l(\Psi^{T}\mi\si_{2}SE_{4}\Psi-\Psi^{*T}\mi\si_{2}SE_{4}\Psi^{*}\r),
\end{equation}
where the symplectic matrix
\begin{equation}
	E_{4}=\l(\begin{matrix}
    0 & \tau_{0}\\
    -\tau_{0}& 0
\end{matrix}\r).
\end{equation}

For $\mr{SU}(4)$ group, it has $10$ symmetric and $5$ anti-symmetric generators. Written in the block representation, these generators are~\cite{PhysRevD.78.115010,Cacciapaglia2014}
\begin{equation}
	S_{a}=\frac{1}{2\sqrt{2}}\l(\begin{matrix}
    \tau_{a}&0\\
    0&-\tau_{a}^{T}
\end{matrix}\r),\quad \text{for  } a=0,1,2,3;
\end{equation}
\begin{equation}
	S_{a}=\frac{1}{2\sqrt{2}}\l(\begin{matrix}
    0& B_{a}\\
    B_{a}^{\dagger}&0
\end{matrix}\r),\quad \text{for  } a=4,...,9
\end{equation}
 with $B_{(4,5)}=\mi^{(0,1)}\tau_{0}, B_{(6,7)}=\mi^{(0,1)}\tau_{3}$ and $B_{(8,9)}=\mi^{(0,1)}\tau_{1}$ for symmetric ones. The remaining five anti-symmetric generators are:
\begin{equation}
	X_{i}=\frac{1}{2\sqrt{2}}\l(\begin{matrix}
    \tau_{i}&0\\
    0&\tau_{i}^{T}
\end{matrix}\r),\quad \text{for } i=1,2,3;
\end{equation}
\begin{equation}
	X_{i}=\frac{1}{2\sqrt{2}}\l(\begin{matrix}
    0& D_{i}\\
    D_{i}^{\dagger}&0
\end{matrix}\r),\quad \text{for } i=4,5
\end{equation}
with $D_{(4,5)}=\mi^{(0,1)}\tau_{2}$. 

The $\mr{SU}(4)$ explicitly (dynamically) reduces to symplectic group $\mr{Sp}(2)$ by the current (constituent) quark mass. The generators of $\mr{Sp}(2)$ obey the relation $S_{a}^{T}E_{4}+E_{4}S_{a}=0$ with mass operator $E_{4}$~\cite{PhysRevD.78.115010,Cacciapaglia2014}. It clearly indicates that the bare mass term is only invariant under the subgroup of $\mr{Sp}(2)$. Thus, $E_{4}$ and $X_{i}$, the elements of coset subgroup $\mr{SU}(4)/\mr{Sp}(2)$, form a six-dimensional vector of quark bilinears ($\Psi^{T}\vec{\phi}\,\Psi+h.c.$) of chiral condensates. $5$ almost Goldstone bosons of $\vec{\pi},\De$ and $\De^{*}$ are excited. Note here that $\vec{\phi}=(E_{4},\mi E_{4}X_{i})$ for $i=1,...,5$ with scalar meson $\si$, pseudo-scalar triplet pions $\vec{\pi}$, scalar diquark and anti-diquark of $\De$ and $\De^{*}$.

Applied the same representation, we rewrite the conventional baryon Lagrangian term $\mu\bar{\psi}\ga_{0}\psi$ into $\mr{SU}(4)$ space as $\mu\Psi^{\dagger}B_{0}\Psi$, where
\begin{equation}
	B_{0}=-\ga_{0}E_{4}=\l(\begin{matrix}
    \tau_{0}& 0\\
   0& -\tau_{0}
\end{matrix}\r).
\end{equation}
Similarly, the isospin density, $\nu\,\bar{\psi}\ga_{0}\tau_{3}\psi$, is written as $\nu\,\Psi^{\dagger}I_{0}\Psi$ with $I_{0}=\mr{Diag}(\tau_{3},-\tau_{3})$ and the axial isospin one, $\nu_{5}\bar{\psi}\ga_{0}\ga_{5}\tau_{3}\psi$, is expressed as $\nu_{5}\Psi^{\dagger}I_{5}\Psi$ with $I_{5}=\ga_{5}I_{0}=\mr{Diag}(\tau_{3},\tau_{3})$.

At finite isospin baryon density, applying the gauge transformation~\cite{KOGUT2000477,PhysRevD.64.016003}
\begin{equation}
	\Psi\to V\,\Psi, \quad V=\exp\l(\mi\th^{i}X_{i}\r)\quad \mr{for}~X_{i}\in \mr{SU}(4)/\mr{Sp}(2),
\end{equation}
one has $\l[X_{i},I_{0}\r]\neq 0$ for $i=1,2$. It draws up that the isospin baryon current is embedded in the charged mesons, $\pi_{\pm}$, to compensate the gauge transformation as it has been well known. Backing from the extended $\mr{SU}(4)$ group to the ordinary $\mr{SU}(2)$ flavor space and replacing the derivatives by covariant one, the Lagrangian term of leading order in chiral perturbation theory ($\chi\mr{PT}$) is shown as~\cite{KOGUT2000477,PhysRevD.64.016003}
\begin{equation}
	\mc{L}_{\chi\mr{PT}}=\frac{f_{\pi}^{2}}{4}\Tr\l(D_{\mu}\Si\r)^{\dagger}\l(D_{\mu}\Si\r)-c\Tr{\l(\Si^{\dagger}+\Si\r)}.
\end{equation}
where unitary matrix $\Si$ is the fluctuations of the order parameters, $\psi^{\dagger}\psi$ or $\psi^{T}\psi$, of symmetry breaking with respect to the extended flavor group. The explicit form of covariant derivative is determined by the constitutes of the order parameter, presented below. The second term is induced by the quark mass.

Furthermore, ignoring the kinetic and mass terms, the remaining static Lagrangian is
\begin{equation}
	\mc{L}_{\chi\mr{PT}}=\mc{L}_{\pi_{i}}\sim \Tr\big[\tau_{3}\nu,\Si\big]\big[\Si^{\dagger},-\tau_{3}\nu\big]\sim\nu^{2}\Tr\l(\tau_{3}\Si\tau_{3}\Si^{\dagger}\r),
\end{equation}
where the commutation bracket is applied because of the bilinear meson field $\psi^{\dagger}\psi$. Unimodular matrix $\Si$ is $2\times 2$ constructing by all the anti-symmetric generators. When the fluctuations of $\Si$ on its Vacuum Expectation Value (VEV) are neglected, the above term becomes {\it negative} for $\Si\sim\tau_{1,2}$. It is the reason why the chiral condensation orient to charged pions at finite isospin baryon density~\cite{PhysRevD.64.016003}.

At finite baryon chemical potential $\mu$, one has $\l[X_{i},B_{0}\r]$ becoming nonzero for $i=4,5$, i.e., diquark channel. Therefore, the covariant form enter into the scalar diquark field, $\psi^{T}C\ga_{5}\psi$, and the effective Lagrangian term behaves as~\cite{KOGUT2000477}
\begin{equation}
	\mc{L}_{\chi\mr{PT}}=\mc{L}_{\De,\De^{*}}\sim \Tr\big\{\tau_{0}\mu,\Si\big\}\big\{\Si^{\dagger},-\tau_{0}\mu\big\}\sim-\mu^{2}\Tr\l(\Si\Si^{\dagger}\r).
\end{equation}
Instead, here anti-commutator rule is used because $\De$ and $\De^{*}$ are composed by two (anti)quarks.

While axial isospin chemical potential $\nu_{5}\neq 0$, we have single generator $X_{3}$ commutes with $I_{5}$
and all other antisymmetric generators have been involved to compensate the gauge transformation of axial isospin current. The associated Lagrangian terms in mesons space are
\begin{equation}\label{eqn_xpt_pi_nu5}
    \mc{L}_{\chi\mr{PT}}=\mc{L}_{\pi_{i}}\sim\Tr\big[\tau_{3}\nu_{5},\Si\big]\big[\Si^{\dagger},\tau_{3}\nu_{5}\big]\sim-\nu_{5}^{2}\Tr\l(\tau_{3}\Si\tau_{3}\Si^{\dagger}\r)
\end{equation}
which is {\it positive} and increases the effective potential for $\Si\sim\tau_{1,2}$. It is the reason why we choose $\pi_{3}$ characterizing the meson type of chiral condensate at finite $\nu_{5}$ in this work. 

Moreover, in the high baryon density limit, when spontaneously flavor symmetric diquark condensates exist, the remaining $\mr{Sp}(2)$ group will keep breaking to a lower symmetry. It is expected that $\mr{Sp}(2)$ will reduce to $\mr{O}(4)$~\cite{PhysRevD.78.115010,Cacciapaglia2014}. We find out that $\l[S_{a},I_{5}\r]\neq 0$ for $a=1,2,8,9$. $\mr{O}(4)$ group is left out due to the six unchanged elements of $S_{a}$. The number of would be Goldstone bosons increases from $5$ to $9$ within two flavors. Two of increased Goldstone bosons are flavor triplet scalar meson, $a_{0}^{\pm}$. The other two are pseudo diquarks, $d,\,d^{*}$. We are not planning to investigating the possible of the condensation of new appearing mesons. The first reason is that $\vec{a}_{0}$ mesons degenerate with $\vec{\pi}$ after $\mr{U}(1)_{A}$ symmetry fully restored. Secondly, our work is exploring in the region of cold high density quark matter where diquark paring pattern and its related behaviors are more interesting than mesons. Coming back, the static Lagrangian term of diquark is modified to %The exploration along the way of $a_{0}^{\pm}$ was discussed in the reference.
\begin{equation}
	\mc{L}_{\chi\mr{PT}}=\mc{L}_{\De,d}\sim \Tr\big\{\tau_{3}\nu_{5},\Si\big\}\big\{\Si^{\dagger},\tau_{3}\nu_{5}\big\}\sim\nu_{5}^{2}\Tr\l(\tau_{3}\Si\tau_{3}\Si^{\dagger}\r),
\end{equation}
which is {\it negative} if $\Si$ originate in the elements of $X_{4,5}$ or $S_{8,9}$. The early two are corresponding to the scalar diquark $\De$ and anti-diquark $\De^{*}$ while as the latter two are axial vector diquark $d$ and anti-diquark $d^{*}$. In two-color space, the diquark, i.e. baryon, is composed as anti-symmetric color singlet. Thus, the flavor symmetric diquarks $d$ and $d^{*}$ have to be symmetric in spin space representing as a spin one state, $(C\ga_{3}\tau_{1})$, to maintain the antisymmetry~\cite{PhysRevD.62.094007,PhysRevD.67.054018,PhysRevLett.90.182002,PhysRevLett.91.242301,PhysRevD.72.034008,PhysRevD.75.054022,PhysRevD.82.085007}. We remind the reader that the choice of $\tau_{1}$ flavor space is confirmed by the generator of $S_{8,9}$. Hence, we have classified that there three kinds of condensates, $\pi_{3},\De\,(\De^{*}), d\,(d^{*})$, mostly take place in at both finite chemical and axial isospin chemical potential from the leading order of low energy effective theory.

% ======================================================================
\section{Numerical Results} 
\label{omega}
Confining in the antisymmetric color sector, for two flavors, a diquark either is an spin, isospin singlet, or an spin, isospin triplet. However, if three colors are available, a competing pattern is to lock the colors to the spin, a linear combination of color structure, where $\la_{A=2,5,7}$ correlate with the spatial direction of $(C\ga_{i=1,2,3})$. It remains an unbroken global $\mr{SO}(3)$ of mixture and the gap is isotropic, which lowing the free energy as shown in the works of~\cite{PhysRevD.72.034008}.

Here we utilize the single Dirac and color component $(C\ga_{3}t_{2})$ since $N_{c}=2$. In what follows we will assume the real quantities of $\De=\De^{*}$, $d=d^{*}$ as usual to apply the mean-field (Hartree) approximation. It is convenient to form the Lagrangian in the Nambu-Gorkov spinors space as what we did in section \ref{qc2d}. The thermodynamical potential is derived as
\bea\label{eqn_om_fir}
	\Om(T,\mu,\nu_{5})&=&\frac{T}{2V}\ln\det\mc{S}^{-1}\l(\mi\om_{n},p;\pi_{3},\De,d\r)+\frac{\pi_{3}^{2}+\De^{2}+d^{2}}{4G}
	\nn &=&-T\sum_{n}\int\frac{\dif^{3}p}{(2\pi)^{3}}\frac{1}{2}\Tr\ln\l(\frac{1}{T}\mc{S}^{-1}\l(\mi\om_{n},p;\pi_{3},\De,d\r)\r) +\frac{\pi_{3}^{2}+\De^{2}+d^{2}}{4G},
\eea
where the sum is over fermionic Matsubara frequencies for $\om_{n}=(2n+1)\pi T$. Taking into account its color, flavor and Dirac structure, $S^{-1}$ is tracing over a $16\times 16$ matrix.

Based on above expression, one obtains three gap equations via minimizing the thermodynamic potential with respect to the mean values of the meson and diquark fields~\cite{HUANG2003835,PhysRevD.70.054013}
\begin{equation}
	\frac{\p\Om}{\p\pi_{3}}=0,\quad \frac{\p\Om}{\p\De}=0,\quad \frac{\p\Om}{\p d}=0.
\end{equation}
With the identity $\Tr\ln A=\ln\det A$, the trace in \Eqn{eqn_om_fir} is evaluated as
\begin{equation}
	\frac{1}{2}\Tr\ln\l(\frac{1}{T}\mc{S}^{-1}\l(\mi\om_{n},p;\pi_{3},\De,d\r)\r)=\sum_{i=1}^{8}\ln\l(\frac{\om_{n}^{2}+E_{i}^{2}}{T^{2}}\r).
\end{equation}
In chiral limit and assuming $\pi_{3}=\De=0$, the double degenerated energy dispersion of quasi-particles is 
\begin{equation}
	E_{i}(p)=\pm\sqrt{d^2+\mb{p}^{2}+\l(\mu\pm\nu_{5}\r)^{2}\pm 2\sqrt{d^{2}p_{z}^{2}+\mb{p}^{2}\l(\mu\pm\nu_{5}\r)^{2}}}.
\end{equation}
$E_{i}(p)$ are guaranteed to be real since $\mb{p}^{2}\geq p_{z}^{2}$ and same as the results in early work without $\nu_{5}$~\cite{PhysRevD.72.034008}. Unfortunately, within nonzero quark mass, or pion superfluid, or including scalar diquark condensation, the energy roots are very complicated, which is not necessary to write it down in analytically but will numerically run in the later.

\begin{figure}[!htb]
\centering
\includegraphics[scale=.55]{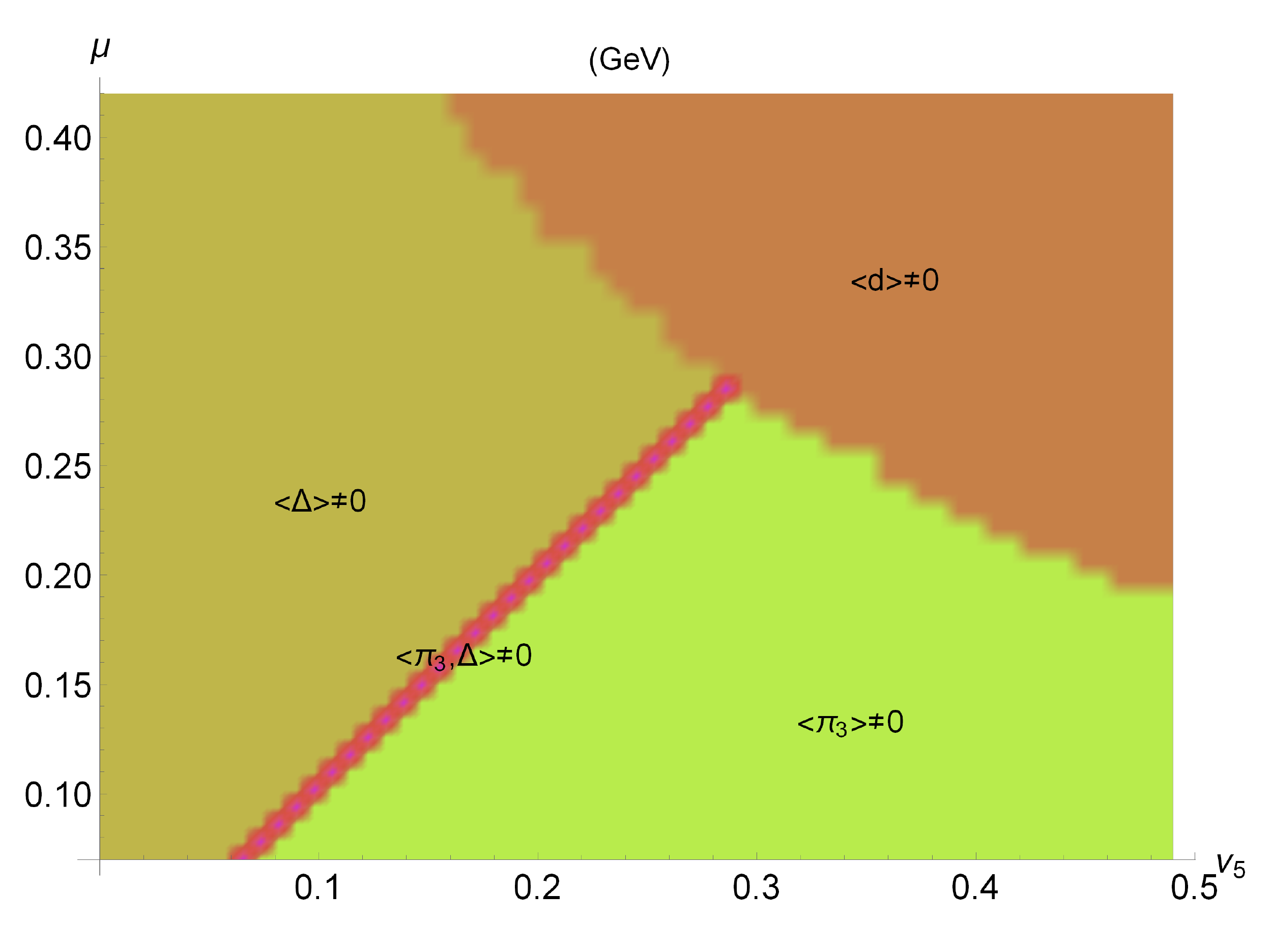}
\caption{Phase diagram of two-color QCD within two fundamental quarks in the $\mu-\nu_{5}$ plane at zero temperature.}
\label{fig_phase}
\end{figure}

Handing above results in and employing the NJL model calculation, the thermodynamical becomes
\begin{equation}\label{eqn_om_sec}
	\Om(T,\mu,\nu_{5})=-\sum_{i=1}^{8}\int\frac{\dif^{3}p}{(2\pi)^{3}}\l[E_{i}(p)\,\th\l(\La^{2}-p^2\r)+2T\ln\l(1+\me^{-\frac{E_{i}(p)}{T}}\r)\r]+\frac{\pi_{3}^{2}+\De^{2}+d^{2}}{4G},
\end{equation}
where $\th$ is the step function. Before proceeding in numerical simulation, we point out that the parameter set is borrowed from Ref.~\cite{PhysRevD.70.054013}, where $\La=0.78~\GeV$, $G=H=10.3~\GeV^{-2}$ and $m_{0}=4.5~\MeV$. Combining the numerical results of $E_{i}$ with \Eqn{eqn_om_sec}, we investigate the phase diagram of two-color QCD with two fundamental quarks at zero temperature, particularly looking into the nature of diquarks. As known before, $\De$ and $\vec{\pi}$ locate in the same coset space of $\mr{SU}(4)/\mr{Sp}(2)$, which means that the mass of $\De$ is same as pions in the vacuum. Without $\nu_{5}$, the diquark condensate onsets while the chemical potential $\mu$ exceeds half of $m_{\De}=m_{\pi}\sim 0.14~\GeV$. Therefore, to focus on the behaviors of diquark, the presenting phase structure is starting from $0.07~\GeV$ for $\mu$. The whole results are plotted in the plane of $\mu-\nu_{5}$, read from \Fig{fig_phase}. One observes it contains four phases, including a mixed state, pion superfluid ($\pi_{3}$), scalar ($\De$) and axial vector ($d$) diquark condensates. The mixed state hybridizes the $\pi_{3}$ and $\De$. It is easily to be understood since both of them characterize the chiral symmetry breaking like mass term, which is the feature of the $\mr{QC_{2}D}$ theory and differs with the reality QCD. Meanwhile, nonzero VEV of $d$ appears at high limit of $\mu$ and $\nu_{5}$. It stems from the symmetry breaking of $\mr{Sp}(2)$ group, rising as a competitor of the chiral condensates, and hence never blending with either $\pi_{3}$ or $\De$.

\begin{figure}[!htb]
\begin{minipage}{0.48\textwidth}
    \centering
\includegraphics[scale=.3]{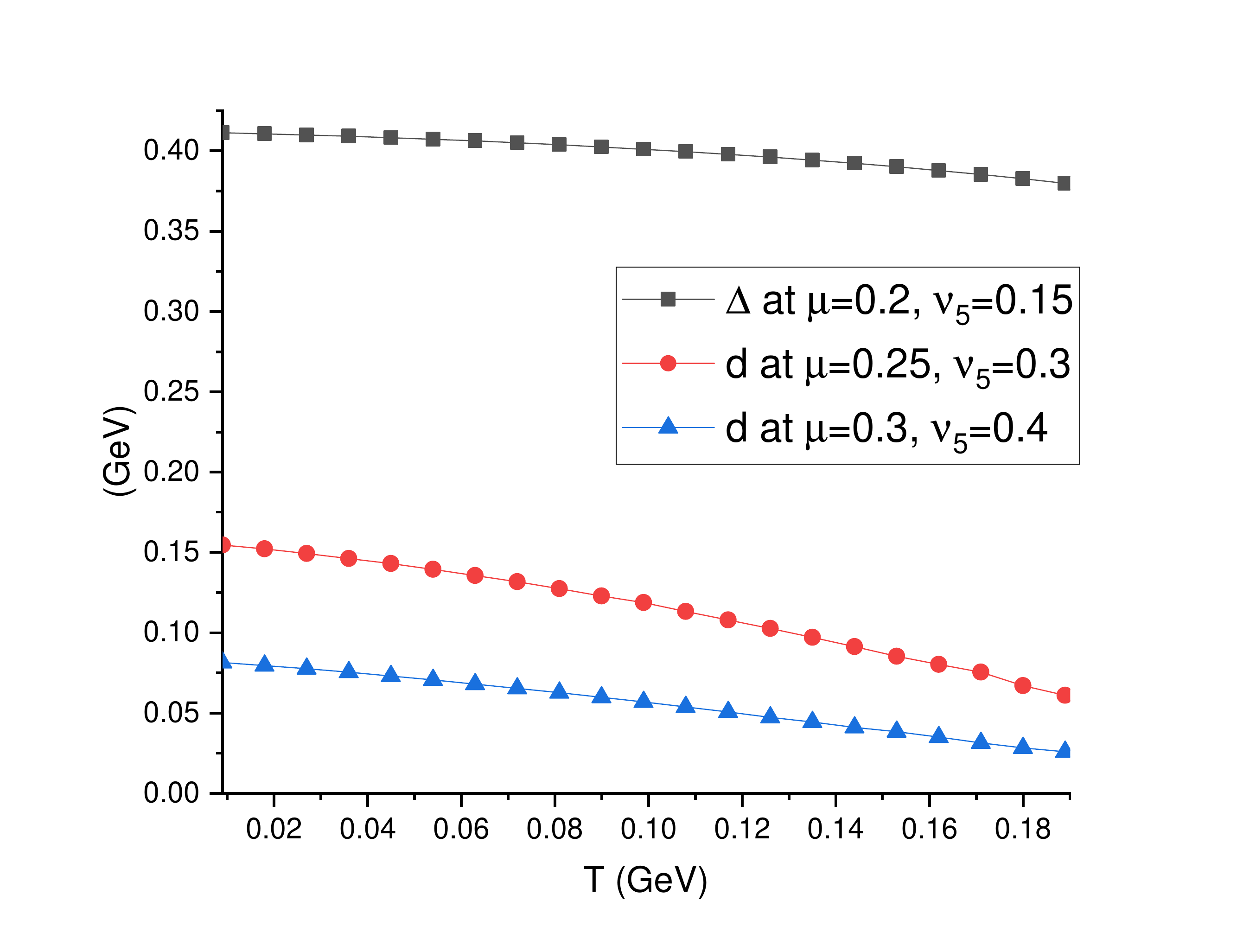}
\caption{Scalar diquark condensate $\De$ and axial vector diquark $d$ as a function of $T$ at given $\mu$ and $\nu_{5}$. The units of $\mu$ and $\nu_{5}$ are $\GeV$.}
\label{fig_res2}
\end{minipage}\hfill
% ======================================================================
\begin {minipage}{0.48\textwidth}
    \centering
\includegraphics[scale=.3]{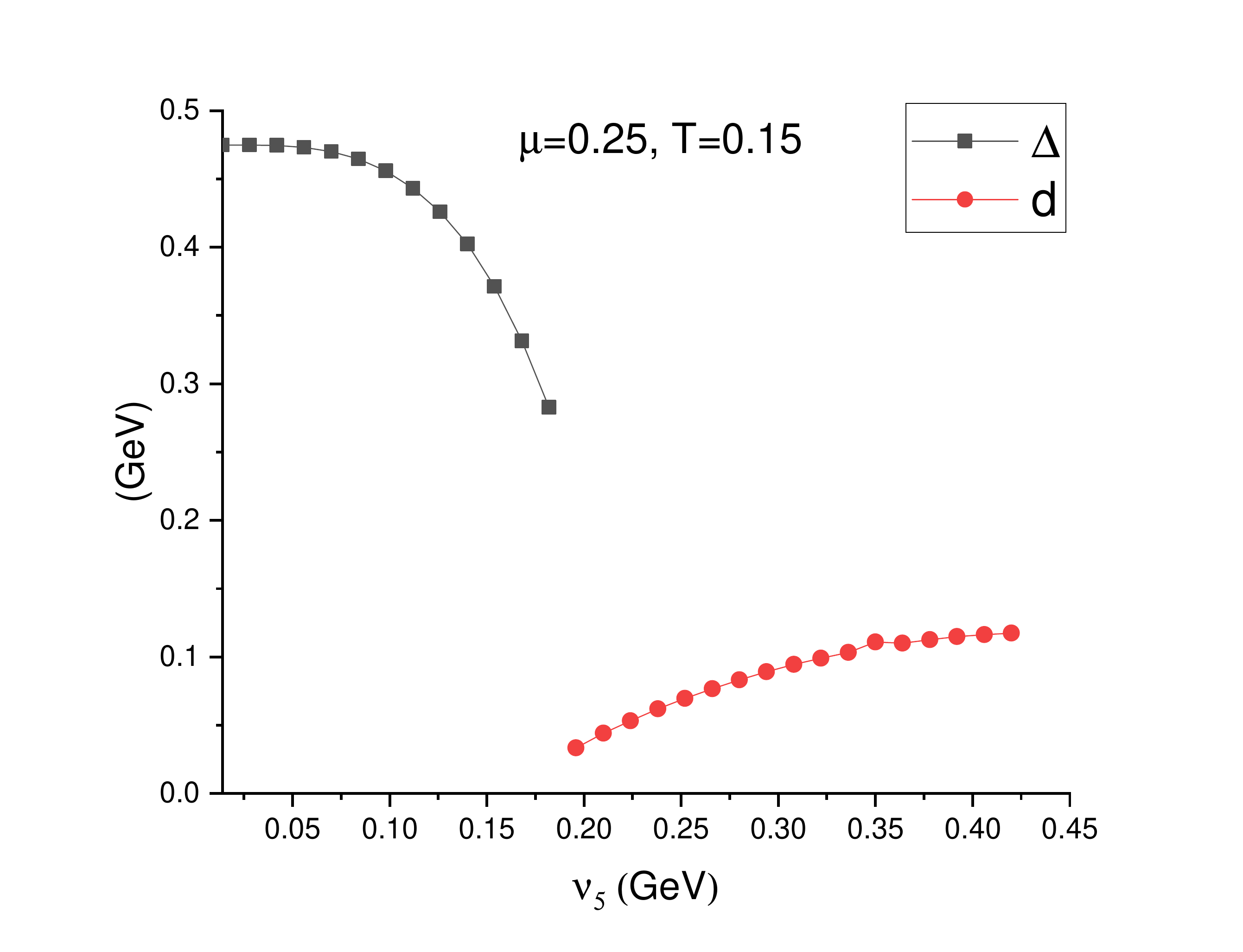}
\caption{Scalar diquark condensate $\De$ and axial vector diquark $d$ as a function of $\nu_{5}$ at given $\mu$ and $T$. The units of $\mu$ and $T$ are $\GeV$.}
\label{fig_res3}
\end{minipage}
\end{figure}

\begin{figure}[!htb]
\centering
\includegraphics[scale=.3]{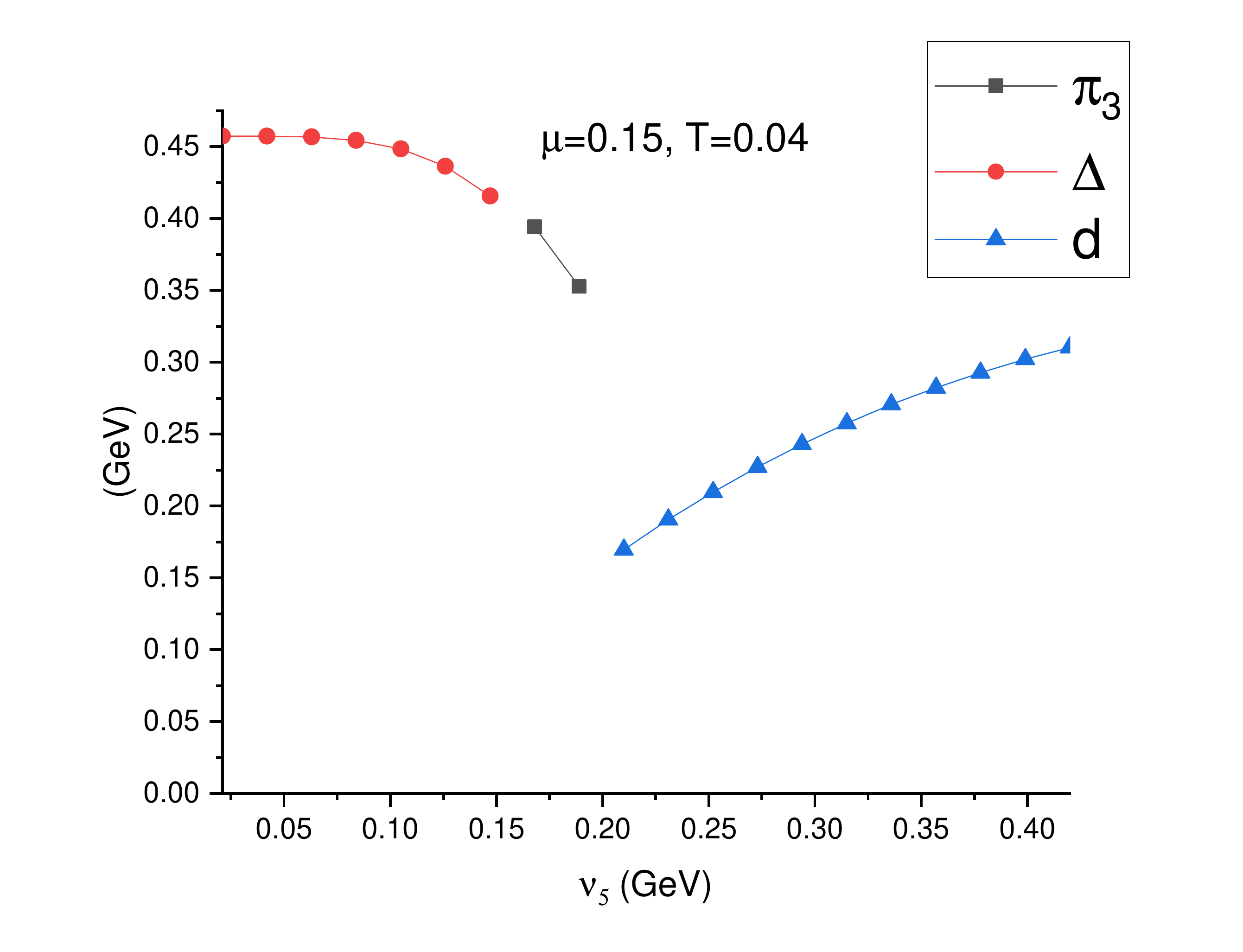}
\includegraphics[scale=.3]{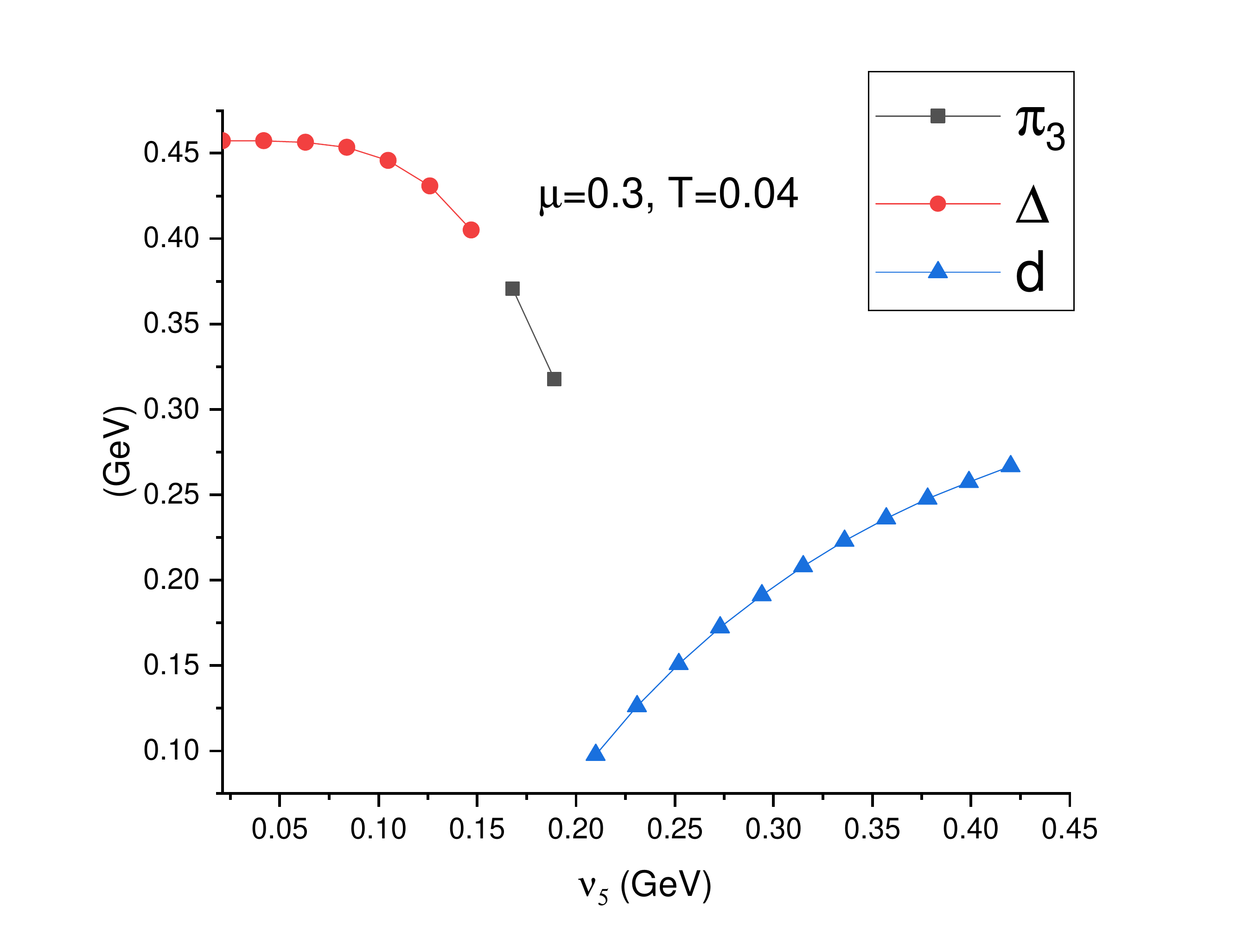}
\caption{Neutral pion condensate $\pi_{3}$, scalar diquark condensate $\De$ and axial vector diquark $d$ as a function of $\nu_{5}$ at given $\mu$ and $T$. The units of $\mu$ and $T$ are $\GeV$.}
\label{fig_res4}
\end{figure}

Before closing this section we mention that the VEV of three condensations are simulated as a function of the axial isospin chemical potential and/or temperatures in \Fig{fig_res2}, (\ref{fig_res3}) and (\ref{fig_res4}). All these plotting demonstrate that the gap of spin one color superconductor is smaller than the chiral condensates, as revealed in the early studies in $N_{c}=3$ color superconductivity~\cite{RevModPhys.80.1455,PhysRevD.72.034008}. We observe that, as temperature increasing, the system undergo a phase transition to quark-gluon plasma at critical temperature, i.e., all the VEV of fields are decreasing, drawn in \Fig{fig_res2}. We show these order parameters as a function of $\nu_{5}$ at different values of chemical potential and temperatures. We figure out that pion superfluid vanishes at high baryon density and leave scalar diquark condensate of $\De$ to identity the breaking of chiral symmetry, presented by \Fig{fig_res3}. The value of the chiral condensate, sum of $\pi_{3}$ and $\De$, is continued before the appearance of the axial vector diquark condensation of $d$, shown in \Fig{fig_res4}.
% ======================================================================
\section{Conclusions} 
\label{con}
In this paper, to mimic the (anti)parallel configured EM field, an axial isospin density term $\nu_{5}$ is proposed to included into the Lagrangian~\cite{PhysRev.177.2426,Bell1969,PhysRevD.63.076010}. The underlying mechanism is because the asymmetric electric-charges of $u, d$ quarks, which inducing an intensive electromagnetic triangle anomaly if the EM field is at the strength of hadron scale. We are mostly interested in the patterns of quark-quark paring since this research launches at finite baryon chemical potential $\mu$ system, starting from $70~\MeV$. To fulfill the capability of comparing with the lattice approach in the future, we employ the two-color QCD toy model to guarantee the positiveness of the Euclidean path integral measurement with pseudo-real fermion sector~\cite{PESKIN1980197,KOGUT2000477,PhysRevD.64.016003}. In this QCD-like theory, the NJL-type interactions are constructed into an extended global $\mr{SU}(4)$ symmetry.

Following the previous works of anisotropic dense quark matter, the chiral rotation in $\mr{SU}(4)$ space is studied by the low energy effective theory, the $\chi\mr{PT}$~\cite{PESKIN1980197,KOGUT2000477,PhysRevD.64.016003}, at leading order, is expressed in Sec.\ref{su4}. This method is reliable even at three color QCD. Its early predictions were confirmed by lattice simulation of two color QCD at nonzero baryon and isospin densities~\cite{PhysRevD.66.014508,HANDS1999327}. We present a full classification of possible QCD vacuum directions determined by the covariance requirement of gauge symmetry. A qualitative difference is figure out between the dense baryon systems of finite isospin and axial isospin, which stems from the different nature of quasi-particles between the mesons and baryons (diquarks). We analytically deliver the alignments of ground state to the scalar diquark, $\De\,(\De^{*})$, or axial-vector diquark, $d\,(d^{*})$, or neutral pion meson directions from the ordinary scalar meson $\si$ condensate in the plane of $\mu-\nu_{5}$. The possible state of pion superfluid agrees with the discovery in the work of \cite{CAO20161}, where the chiral rotating to $\pi_{0}$ is driven by the electromagnetic anomaly. We emphasize that $\pi_{0}$ and $\pi_{\pm}$ lost their degeneracy in the EM fields. The system is suggested to turn toward the condensation of charged pion instead of neutral one if more processes of EM anomaly included, such as five-point contribution of $\ga\ga\to\pi_{0}\pi_{+}\pi_{-}$~\cite{PhysRevD.4.3497,1971JETPL1494T} in the work of~\cite{Chao:2018ejd}. Therefore, there is no discrepancy to achieve the result in \Eqn{eqn_xpt_pi_nu5} and the charged pion condensation is absent for nonzero $\nu_{5}$ since axial isospin current is generated by the triangle diagram alone.

Searching along these three given orientations, applying the mean-field approximation, we then numerically investigate the phase structure by the two-color NJL model in Sec. \ref{omega}. To our knowledge, the final model calculation is consistent with the expectation of $\chi\mr{PT}$. We stress that a novel result is revealed in the QCD phase diagram, where a co-existent phase appears. It mixes the neutral pion and scalar diquark condensates in a shallow window where $\mu$ being proportional to $\nu_{5}$. The relevant thermodynamic quantities run at nonzero temperature, as well.

We remark that in this simplified model, the back reaction on the electric-blind fields hasn't been taken into account. The coupling of the quark sector to the Polyakov loop is not enclosed and therefore it is not able to yield the (de)confinement effect~\cite{PhysRevD.88.025016,PhysRevD.91.125040}. Furthermore, a first principle calculation can be pursued by Dyson-Schwinger equations and/or functional renormalization group approaches. Nevertheless, our study is devoted to the properties of axial isotopically asymmetric dense QCD matter. It can be verified by lattice QCD simulation via applying similar strategies at both finite baryon and isospin chemical potential. We wish the success of the present probing call an attention in the lattice community.

Application of our analysis to heavy-ion colliding systems, where strong anti-parallel electric and magnetic field configuration is produced above and below the reaction plane, is speculated. In addition,, we observe that axial vector diquark condensate phase shows up as a physical ground state in the phase diagram. The existence of this phase implements the purpose of low-temperature/high-density ion-collision experiments at different worldwide facilities and supports a further investigation to compact stars.

In sum, our research riches the phenomenology of QCD-like matter under different extreme conditions. The study of two-color QCD serves as an first step towards to the reality QCD matter. It will be our upcoming topic to extend the exploration to $N_{c}=3$ QCD matter. We believe that our future efforts will contribute to a general understanding of cold dense strongly interacting matter.
% ======================================================================
% ======================================================================
% ======================================================================
% ======================================================================
% ======================================================================
\section{Acknowledgments}
\label{Ackn}
I thank M. Huang, K. Xu and M. Ruggieri for discussions and the comments from T. Schaefer. This work is supported by the NSFC under Grant number: 11605254.
% ======================================================================

%\newpage
%\bibliography{diquark} 
%\bibliographystyle{unsrt}
\end{document}